\documentclass[doublecol,12pt]{iopart}

\usepackage{graphicx}
\usepackage{color}
\usepackage{dcolumn}
\usepackage{bm}

\begin{document}

\title[]{Coherent Transport and Symmetry Breaking - Laser Dynamics
  of Constrained Granular Matter}

\author{Andreas Lubatsch$^{1}$ and Regine Frank$^{*2,3}$}

\address{
$^{1}$Electrical Engineering, Precision Engineering, Information
  Technology,\\
{\color{white} $^{1}$}Georg-Simon-Ohm University of Applied Sciences,\\
{\color{white} $^{1}$}Kesslerplatz 12, 90489 N\"urnberg, Germany\\                 
$^{2}$Institute of Theoretical Physics, Optics and Photonics,\\
{\color{white} $^{1}$}Center for Collective Quantum Phenomena (CQ) and\\ 
{\color{white} $^{1}$}Center for Light-Matter-Interaction Sensors and
Analytics (LISA+),\\
{\color{white} $^{1}$}Auf der Morgenstelle 14, Eberhard-Karls-Universit\"at, 72076 T\"ubingen, Germany\\
$^{3}$Institute of Solid State Physics, Wolfgang-Gaede Strasse 1,\\
{\color{white} $^{1}$}Karlsruhe Institute of Technology (KIT), 76131
Karlsruhe, Germany
}
\ead{r.frank@uni-tuebingen.de}
\date{\today}

\begin{abstract}
We present diagrammatic transport
theory including self-consistent nonlinear enhancement and dissipation in the
multiple scattering regime. Our model of Vollhardt-W\"olfle transport of
photons is fit-parameter-free and raises the claim that the results hold up to
the closest packed volume of randomly arranged ZnO Mie scatterers. We find that a symmetry breaking caused by
dissipative effects of a lossy underlying substrate leads to qualitatively
different physics of coherence and lasing in granular amplifying media. According to our results, confined and
extended mode and their laser thresholds can be clearly attributed to unbroken
and broken spatial symmetry. The diameters and emission profiles of random
laser modes, as well as their
thresholds and the
positional-dependent degree of coherence can be checked experimentally.

\end{abstract}

\maketitle
\section{Introduction}
After two decades of random laser research
\cite{Cao,Vaneste01,Mujumdar04,Vaneste07,Muskens,Jaq} these systems are still
highly fascinating. 
The involved rich physics and also possible applications
\cite{Stone,Leonetti} in systems that, at first glance appear to be just
'dust powder' have started to reveal more and more fascinating details. 
Absolutely essential for the fundamental behavior of a random laser
is the spatial extension of random lasing modes. If the lasing spots are
strongly confined, the random laser actually is operated as a collection of single-mode
lasers where the modes do not overlap in space \cite{Mosk_Mode,Vardeny}. On the other
hand, the random laser can exhibit another type of mode which covers the whole
ensemble showing significantly different emission characteristics and a higher
laser threshold. Additionally all these properties are derived without
confining external resonator and the modes seem to be only due to transport
through disordered granular media and amplification. The experimental finding that spatially confined and
  extended laser modes can actually co-exist in the same region of strongly
  scattering nano-crystalline powders \cite{Kalt09,Kalt10} has been completely
  counterintuitive. Nevertheless an 'ab
  initio' description for coherent emitting modes in diffusive, weakly or
  strongly localizing systems could not yet be given
  for this phenomenon. In this paper we derive by means of quantum field
  diagrammatical photon transport incorporating several loss channels spatially confined and extended random
  laser modes which may co-exist. It is proven that
  the experimentally observed mode types in different gain regimes can be
  explained in a single framework of transport renormalized by
  dissipation. Dissipation processes are not only frequency selective with
  respect to the absorption and transmission properties of the substrate, they
  can be further
  influenced by the dispersity of the 
  powder, and the nonlinear enhancement. We show that the emission statistics,
  the coherence and the threshold of random laser modes are severely changing
  due to symmetry breaking of photonic transport by dissipation. However we
  find that also modes with strong losses arrive at a laser threshold. This result can be
  checked by measuring the extent and the degree of coherence of
  random laser modes relying on non-symmetric boundaries.  

\section{Light in Granular Matter}
\begin{center}
\begin{figure}[t!]
\qquad\qquad\qquad\begin{center}\vspace*{0cm} \hspace*{0cm}\rotatebox{0}{\scalebox{0.32}{\includegraphics{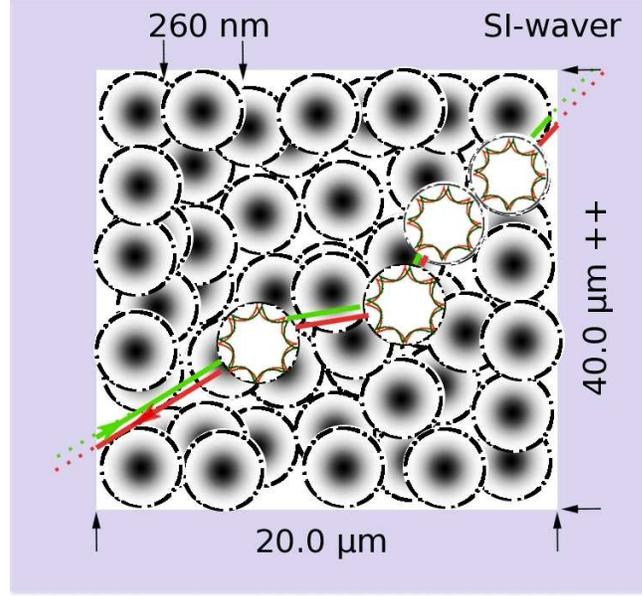}}}
\vspace*{0.0cm}
\caption{Schematic setup of ZnO nano grains embedded in a Si-waver. The spheres' diameter is $d=260\,nm$. The samples' size is assumed to be of $20.0\,\mu m\, \times\,
  40.0\,\mu m ++$. The
  non-quadratic size is assumed to show the symmetry breaking due to a high
  aspect ratio and lossy Si boundary. The sample extent is $4.0\,\mu m$ in the third
  dimension. The volume filling is $50 \%$. Far below the laser
  threshold the permittivity of ZnO is given by $Re\,{\epsilon_s=4.0164}$. Red
  and green paths represent counter propagating, time reversal, photons coupled
  in non-linear response of ZnO and the Mie-resonance of the wave.}
\label{exp}\end{center}
\end{figure}
\end{center}

Light transport in random media is a very fascinating subject. In non-linear
granular systems of low packing radiation transport obeys the well known
diffusion equation, whereas in densely packed random media coherent transport
sets in which is formost seen in a deviation from the exponential decay of
diffusive light intensity, the long-time tail. Fancy effects like coherent
backscattering (CBS), a factor of two of light intensity in the exact backscattering
direction, can be observed. Diagrammatic transport of light intensity which treats light propagation as
well defined paths of photons (see sketch
Fig.\,\ref{exp}) can describe these observations. A photon (green line) is
scattered by active particles, possible spheres. This means that it
is absorbed by a particle and excites the electronic structure of the
material as well as the internal geometrical resonance, Rayleigh-scattering or
Mie-scattering. During the (re-)emission process frequency conversion and decoherence processes of the
electronic sub-structure can lead to frequency changes. The exact time
reversal procedure is denoted red. Both photons, the forward propagator and
the time reversal procedure can interfere. A perfect interference or
correlation the of propagator and time-reversal procedure is represented in
diagrammatics by the most crossed diagram, the Cooperon \cite{Woelfle80,Akkermans}. Cooperons again may
suffer destructive influences at the samples boundaries. If photons
leak out or if their frequency is absorbed, the symmetry is
broken. Absorption processes can efficiently be used to tune light
transport in random systems as well as waveguides \cite{Yamilov}.

Lasing in granular media is often described in terms of random cavities
\cite{Shapiro} modes or
quasi-(leaky) modes \cite{Andreasen}. These modes are by definition coherent and
therefore obey Poissonian statistics. They have been estimated to form so
called random cavities, in other words artificial but random cavities which
are determined by the local scatterer arrangement. This scenario is of course
possible, but fundamentally different from what we describe in this letter.

We investigate with the diagrammatic ansatz random lasing in nanocrystalline
ZnO samples embedded in small depressions etched in a Si-waver. A similar
experimental setup can be found in \cite{Kalt09}. It has been found that at first
laser threshold a spatially confined mode starts to lase symmetrically in
space. With increasing pump strength, a rising number of modes of this type
may be found strongly localized at several spatial positions. Increasing the
excitation power further a second laser threshold for
a different lasing frequency of another type of mode is found. The diameter of
this mode is
large compared to the others, it may cover the whole
sample. The physics of both modes is fundamentally different and we will see
that the extended mode arises in principle only for a symmetry breaking due to
dissipation.

\section{Coherent Photon Transport} 
We use a diagrammatic field theory ansatz for light in a diffusive system
including interferences, Vollhardt-W\"olfle theory of photons \cite{Frank11,Lubatsch05} which has proven to be rigorous for signatures of Anderson
localization in non-linear random media \cite{Maret}. 
Vollhardt-W\"olfle ansatz precisely means that the modes we derive here are in their
information value not restricted to the coherence of the wave but they
additionally describe the coherence of transported light intensity and its
decay. As already outlined the difference in the diameters
can not be explained just by the increase of the pump power or the dispersity
of the powder alone. Only investigating the possibility of
frequency dependent dissipation varying in space, the absorption of photons by
the crystal substrate at the boundary, yields the difference of the diameters.
Loss initially suppresses a large number of modes which only arrive at their
lasing threshold eventually for significantly higher pump strengths when the
intrinsic nonlinear gain yields a balanced process. Strictly speaking, several
spatially and spectrally distinctive loss channels within the ZnO
sample and the Si-crystal substrate \cite{Palik} lead to the co-existence of both
types of modes; The extended mode overall loses more intensity. It seems at
first sight that the principle itself is reprising at another intensity scale
which is induced by further degrees of freedom, but the study of the correlation
length with respect to the position in the granular matter will clearly bring
forward that symmetry breaking due to loss causes fundamental differences. One could even imagine tuning
the powder's parameters in such a way that a stepwise access to different loss
mechanisms could be possible and the random laser therefore could be
controlled by the ensemble size, the surface, the type of substrate etc..\\
Theoretically the non-linearity is actually being established
by a doubly nested self-consistent calculation. In the following, a description for correlation and coherence of light in
terms of wave and particle is explained. The photon density
response, the four-point correlator is derived from Bethe-Salpeter equation
(BS) for photons,

\begin{figure*}[t!]
\quad\hspace{0cm}\rotatebox{0}{\scalebox{1.0}{\includegraphics[width=1.0\textwidth]{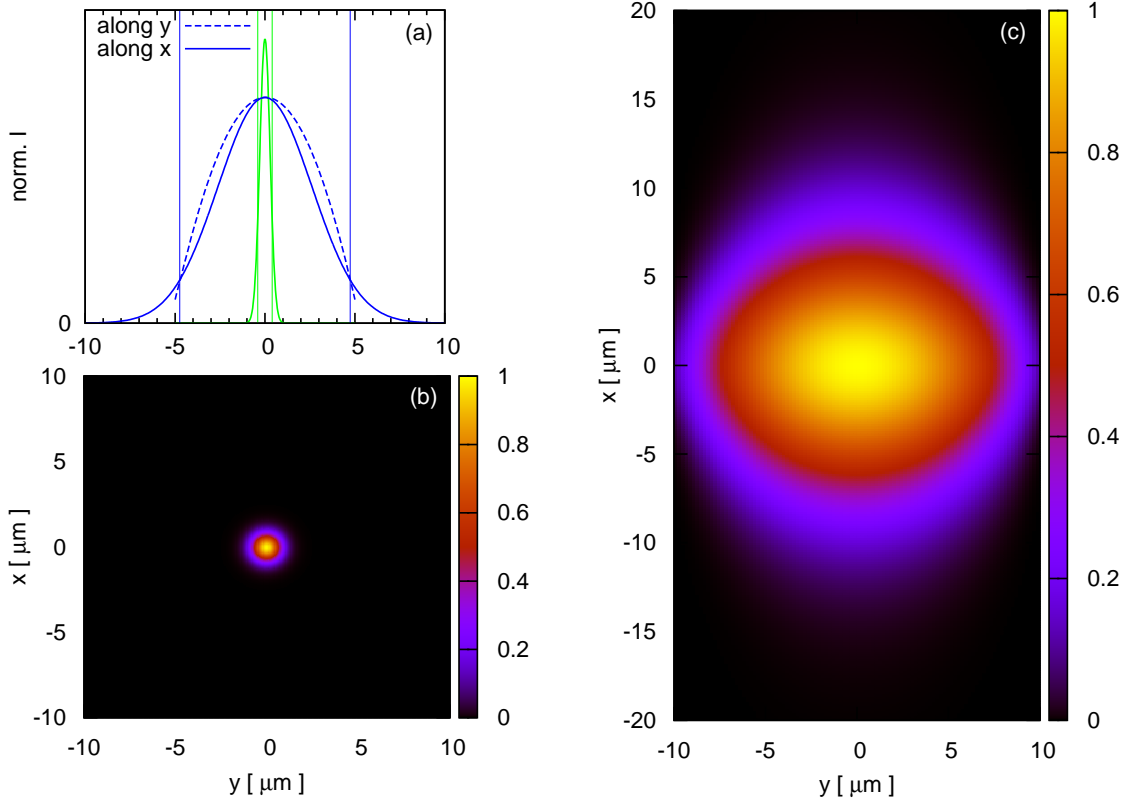}}}
\vspace*{0.0cm}\\
\caption{Computed lasing mode diameters and intensity distribution (color
  bar). Samples parameters can be found in Fig.\,\ref{exp}. Results are shown for
  homogeneous 2 photon pumping $ \lambda = 355$ nm (bulk ZnO
  bandedge). (a) Comparison of the intensity
profile through the mode center for the symmetric confined mode (green), the
transport mean free path is $l_s= 499.2$ nm at the sample center, and the
extended mode, $l_s= 501.57$ nm at the same point, (blue lines, taken along x-
and y-direction). Both modes are shown at the threshold. Mean free paths
differ due to renormalization from lossy boundaries. Vertical lines
represent the decay to $1/e$ compared to the modes maximum intensity. Both
modes are spectrally separated as explained in the text. Corresponding laser
dynamics is shown in Fig.\,\ref{thresh}. (b) Confined mode, intensity distribution. Emission energy is
$3.23 eV$. (c) Extended mode, emission energy $3.21 eV$. The color gradient
in both cases denotes the absolute amount of
coherently emitted lasing intensity with its spacial dependency.}
\label{DUAL}
\end{figure*}

\begin{eqnarray}
\Phi = G^RG^A[1 + \int\frac{d^3q}{(2\pi)^3} \gamma\Phi]
\end{eqnarray}

which reads in position space

\begin{eqnarray}
\label{BS_real_space_neu_01}
\Phi (r_1,r_1';r_2,r_2')&=& G^R (r_1,r_1')G^A
(r_2,r_2')\,+\qquad\qquad\hrulefill\\
\qquad&&\sum_{r_3,r_4,r_5,r_6}\!\!\!\!\!\!\!\!G^R (r_1,r_5) G^A (r_2,r_6)\,\gamma(r_5,r_3;r_6,r_4) \Phi(r_3,r_1';r_4,r_2').\nonumber
\end{eqnarray}

The numbering marks independent positions within space, the dashes denote the
selfconsistency of the diagram.
All
  interference effects are considered by means of the irreducible vertex
  {\bf$\gamma$} which includes most crossed diagrams
  (Cooperons) yielding  memory effects as well as retardation and
  finally cause second order coherent emission of random lasers. From BS a
  Boltzmann equation is derived which yields two independent equations the
  continuity and the current density relation. Local energy conservation is guaranteed by means of a Ward identity
  (WI). 
The sample is modeled by a system which is large (infinitely) sized in one
direction but finite otherwise. The Fourier transform in the infinite
direction $x$ of Eq. (\ref{BS_real_space_neu_01}), and use of the expression for the 
single particle Green's function $G\,=\,[\epsilon_b(\omega/c)^2-|{\bf q}|^2
-\Sigma^{\omega}_{{\bf q}}]^{-1}$,
leads to the kinetic equation for the correlator $\Phi$ with spatial dependencies due to the
additional loss channels at the boundaries of the finite $y$-direction,

\begin{eqnarray}
\!\!\!\!\!\!\!\!\!\!\!\!\!\!\!\!\!\!\!\!\!\!\!\![\Delta \Sigma + 2 {\rm {Re}}\,\epsilon_b \omega\Omega &-& 2 {\rm{Im}}\,\epsilon_b\omega^2 - 2\vec{p}_{x}\cdot \vec{Q}_{X}
+ 2ip_y\partial_Y] \Phi_{pp'}^{Q_{X}} (Y  , Y')\nonumber \\
&=& \Delta G_p (Q_{X}; Y, Y')
\delta(p-p')\nonumber\\
&&\qquad+
\sum_{Y_{3,4}}
\Delta G_p (Q_{X})\,
\int \!\!\!\frac{{\rm d}p'' }{(2\pi)^2}
\gamma_{pp''}^{Q_{X}}  (Y,Y_{3,4})
\Phi_{p''p'}^{Q_{X}}  (Y_{3,4} , Y').
\label{kin_new}
\end{eqnarray}

$\Delta G = G^R -G^A$, $p$, $p'$ and $p''$ are momenta. The scatterer's geometric properties are represented within the Schwinger-Dyson (SD)
  equation {\bf$G=G_0 + G_0TG$} which leads to the solution for the Green's function {\bf$G^R$} and
  {\bf$G^A$} of the
  electromagnetic field, the light wave. Extended amplifying Mie spheres
  \cite{Mie} as
  scattering centers \cite{Lagendijk06} are
  represented by the self-consistent complex valued scattering matrices
  {\bf$T$} leading to the self-energy $\Sigma$, which is
    derived in the independent scatterer approximation here. The ZnO scatterer's initial permittivity is
  given by $Re\,{\epsilon_s=4.0164}$, the imaginary part $Im\,{\epsilon_s}$, the
  microscopic gain, is derived
  self-consistently in what follows, yielding saturation effects. The
  background is air $\epsilon_b=1$. The photon density emitted from the
  amplifying Mie particles is derived by means of coupling to a rate equation
  system (see next section). It is therefore self-consistently connected to
nonlinear gain and the
dielectric function {\bf$\epsilon=\epsilon_L + \epsilon_{NL}$}. The latter yields
finally nonlinear feedback in both, electromagnetic wave transport (SD frame) and intensity
transport (BS frame). 

Dissipation processes at the boundaries severely influence the Green's functions formalism. It is well known that Green's
functions, intended to describe the transport of photons in the random laser
system on the one hand but being a description of
microcanonical ensembles on the other hand, have to obey time reversal
invariance. However within grand canonical (open) ensembles of random lasers the entropy is
increased by photonic intensity transport processes which nevertheless foot on
time reversal symmetry for the propagation of the electromagnetic wave. This aspect of
dissipation and disorder guarantees the completeness of the  'ab
  initio' description of the
propagating light intensity by the four-point correlator {\bf$\Phi =
  A\,\Phi_{\epsilon\epsilon} + B\,\Phi_{J\epsilon}$} here given in terms of the
momenta. {\bf$\Phi_{\epsilon\epsilon}$} equals the energy density and
{\bf$\Phi_{J\epsilon}$} equals the energy current, {\bf$A$} and {\bf$B$} are
pre-factor terms derived in \cite{Frank11}. Starting with the
renormalized scattering mean free path {\bf$l_s$}, the framework yields all
relevant transport lengths and includes all interference effects. The modal behavior,
the core of the random laser, is described efficiently by the determination of
the correlation length {\bf$\xi$} with respect to various spatially dependent loss channels. The co-existence of strongly confined and extended modes can be consistently explained.
BS is solved in a sophisticated regime of real space and momentum with respect
to the high aspect ratio of the random laser sample and the description
for the energy density {\bf$\Phi_{\epsilon\epsilon}(Q,\Omega)$} is derived which is
computed regarding energy conservation

\begin{eqnarray}
\Phi_{\epsilon\epsilon}(Q,\Omega)=
\frac
{N_{\omega}(Y)}
{
\Omega + i D Q_{X}^2
\underbrace{- i D
  {\chi_{d}^{-2}}-{c_{1}\Big(\partial^{2}_{Y}\Phi_{\epsilon\epsilon}(Q,\Omega)\Big)+ c_{2}}   + i D  \zeta^{-2}}  _{i D \xi^{-2}} }.
\label{Pole}
\end{eqnarray}

Here the numerator {\bf$N_\omega$} is basically representing the local density of photonic
modes LDOS which is sensitive to amplification and absorption of the
electromagnetic wave. $Q$ is
the center of mass momentum of the propagator denoted in Wigner coordinates,
the index denotes the Fourier partner. $\Omega$ is the
center of mass frequency and $D$ is the self-consistently derived diffusion
constant, $c_1$, $c_2$ are coefficients explained in \cite{Frank11}. 

\begin{center}
\begin{figure}[t!]
\qquad\qquad\qquad\vspace*{0.2cm} \hspace*{0cm}\rotatebox{0}{\scalebox{0.4}{\includegraphics{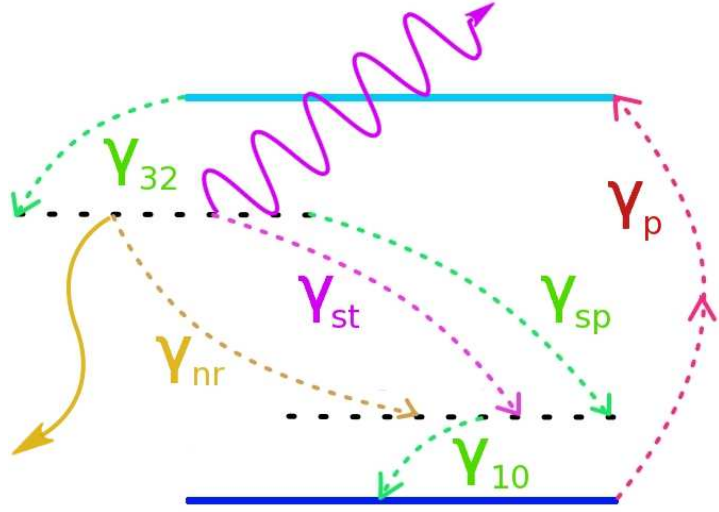}}}
\vspace*{0.0cm}
\caption{Schematic representation of the implemented 4-level laser rate
  equations. The electronic transitions (dashed lines) are due to 2-photon
  excitation $\gamma_p$, spontaneous decay $\gamma_{sp}$, non-radiative decay
  $\gamma_{nr}$, stimulated emission (pink) $\gamma_{st}$ and the
  transitions $\gamma_{32}$ and $\gamma_{10}$ which are necessary to derive a
  threshold behavior. The spontaneous emitted photon is not displayed for
  keeping the clarity of the sketch.}
\label{setup_laser}
\end{figure}
\end{center}

The equation for the energy density $\Phi$, Eq. (\ref{Pole}), represents a new
result describing the propagation of photons within a
finite sized sample. This is in contrast to previous work where the results
were derived solely in momentum space for infinite sized systems.
Here we follow methodologically the systematically outlined
transport scheme in \cite{Frank11} but additionally we take into account 
the spatial dependence as denoted in Eq. (\ref{kin_new}) and displayed in Fig. \ref{exp}:
The
sample shall be homogeneously pumped from above. Diffusive transport, especially interferences occur
preferentially on long paths in-plane of the large scaled random laser
sample. The physics of most crossed diagrams therefore significantly dominates
the coherence properties: Dissipation and losses due to
spontaneous emission and
non-radiative decay are basically homogeneous, however at the samples edges
the situation changes qualitatively. Here transport is inhibited and photons are absorbed within the Si-substrate. This frequency
selective dissipation severely affects the spectrum of lasing modes and their
diameters. All these channels are
represented within the pole of Eq.\,(\ref{Pole}) resulting in separate dissipative
length scales $ \zeta $ due to homogeneous losses, and $\chi_{d}$ due to gain
and absorption that go along with photonic transport and the open or strongly
absorbing Si-boundaries (see Fig.\ref{setup_laser}). The full dissipative influence on
coherent propagating photons and wave is found within the
renormalized so called mass term of the diffusion equation:

\begin{eqnarray}
i D \xi^{-2}=
 - i D
  {\chi_{d}^{-2}}-{c_{1}\Big(\partial^{2}_{Y}\Phi_{\epsilon\epsilon}(Q,\Omega)\Big)+
    c_{2}}   + i D  \zeta^{-2}.
\label{diff}
\end{eqnarray}

By solving of the non-classical diffusion equation
Eq.\,(\ref{diff}) the coefficients $c_1$ and $c_2$ are selfconsistently derived, and we arrive the spatial distribution of
energy density:

\begin{eqnarray}
-\frac{\partial^2}{\partial Y^2}  \Phi_{\epsilon\epsilon}
=
 \frac{1}{D}\left[
\frac{D}{- \chi_d^2} +\frac{D}{\zeta^2 }
\right]
\Phi_{\epsilon\epsilon}
+ {\rm ASE}.
\label{SE}
\end{eqnarray}

The nonlinear self-consistent microscopic random laser gain $\gamma_{21}
n_2$ (see next section) incorporates the influences of both length scales $\chi_d$ and $\zeta$, 

\begin{center}
\begin{eqnarray}
\frac{D}{- \chi_d^2 }+\frac{D}{\zeta^2 }
=
\gamma_{21}  n_2 ,
\end{eqnarray}
\end{center}

and therefore represents the physical properties of the random laser samples
within the absorptive Si-waver. $\gamma_{21}$ is the transition rate of stimulated
emission and $n_2$ equals the selfconsistent occupation of the upper laser
level (see next section). The abbreviation {\bf$\rm ASE$} on the right of Eq.\,(\ref{SE}) represents
all transport terms yielding amplified spontaneous emission.

\section{Lasing  and Threshold Behavior}
The diagrammatic transport approach as such is successful in explaining
strong and Anderson localization. We focus now on the implementation
of a realistic semi-conductor ZnO which is able to perform a laser transition.
In order to describe lasing action, the electronic dynamics have
to be accounted for \cite{John04}. A popular way to do so, is to consider an electronic system
consisting of four energy levels and write down coupled equations for
the occupation numbers of the individual energy levels (see Fig.\,\ref{setup_laser}). The resulting laser
rate equations are given as

\begin{eqnarray}
\frac{d}{dt} n_3(t)&=& \gamma_P -(1/\tau_{32})n_3\label{lrg1}\nonumber\\
\frac{d}{dt}  n_2(t) &=& (1/\tau_{32})n_3(t)- (1/\tau_{sp}) n_2(t)\label{lrg2}\nonumber\\
 &-&(1/\tau_{21})[n_2(t) - n_1(t)]n_{Ph}(t) - (1/\tau_{nr}) n_2(t)\nonumber\\
\frac{d}{dt}  n_1(t) &=& -(1/\tau_{10})n_1(t) + (1/\tau_{sp}) n_2(t)\label{lrg3}\nonumber\\
 &+&(1/\tau_{21})[n_2(t) - n_1(t)]n_{Ph}(t) + (1/\tau_{nr}) n_2(t)\nonumber\\
\frac{d}{dt}  n_0(t)&=&(1/\tau_{10})n_1(t) - \gamma_p\label{lrg4}\nonumber\\
\label{lrg4}
\end{eqnarray}

In the above equations Eqs. (\ref{lrg4}), $\gamma_P$
is the external pump rate, $n_{0-3}$ are electronic populations of the levels respectively,
$\tau_{ij}$ are the states' lifetimes $\frac{1}{\tau_{ij}}=\gamma_{ij}$,
$\tau_{nr}$ is the non-radiative decay time, $\tau_{sp}$ represents the spontaneous
decay time and $\tau_{21}$ is the time scale of the lasing transition. 
The term $[n_2(t) - n_1(t)]n_{Ph}(t)$ marks the inversion of the occupation
numbers of level $1$ and $2$ proportional to the number of stimulated
emitted photons $n_{ph}$. 
The spatial coordinates are suppressed in equations
Eqs. (\ref{lrg4}) for the clarity of presentation.

\begin{center}
\begin{figure}[t!]
\qquad\qquad\qquad\vspace*{0.5cm} \hspace*{0.0cm}\rotatebox{0}{\scalebox{0.4}{\includegraphics{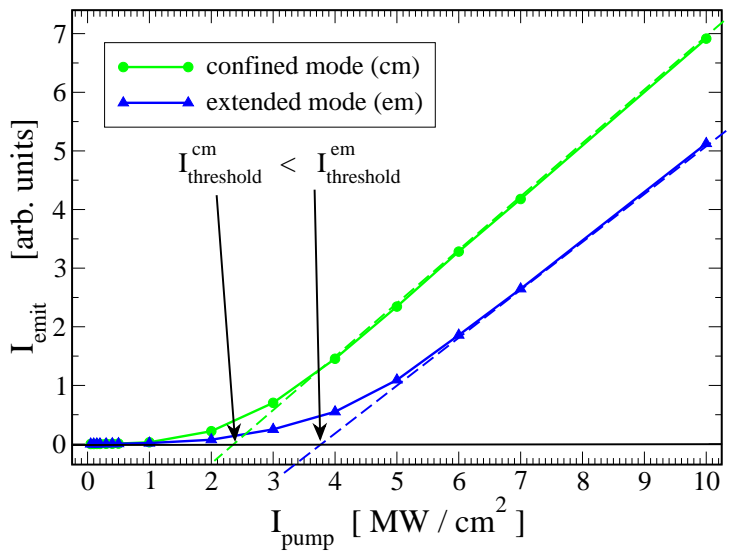}}}
\caption{Laser dynamics and thresholds derived by the solution of the coupled system of
  mesoscopic transport and semiclassical time dependent laser rate equations
  taking into account energy conservation. Extended and confined
  modes ($3.21 eV$ and $3.23 eV$, see Fig.\ref{DUAL}) are suffering different
  types of loss by means of transport and absorption, as well as spontaneous emission
  and nonradiative decay. The extended modes suffer in sum a stronger loss of
  intensity, and therefore arrive at their laser thresholds for significantly
  higher pump intensities.}
\label{thresh}
\end{figure}
\end{center}

The last and most likely the crucial step is to couple the electron dynamics to
the propagating  energy density of the lasing photon density. This is achieved
by identifying the growth of the photonic energy density with a corresponding 
population inversion in the laser rate equation. 
The system is solved also dependent in time and the typical threshold
behavior of the stimulated emitted photon number density is derived 
which matches the experiment for both lasing modes,
confined as well as extended modes (see Fig.\,\ref{thresh}). Assuming $5.0\,ns$
pulses the self-consistent
laser threshold of the confined mode is derived to be $\sim
2.4 MW/cm^2$ and for the extended mode to be $\sim 3.7 MW/cm^2$.

\section{Results and Discussion}

Numerical calculations of self-consistent photonic transport theory and random
lasing lead to a variety of directly measurable quantities. Above all we mention that in disordered amplifying and constrained samples like the Si-confined ZnO powder here, the gain is clearly influenced by the
boundary and consequently dependent to the position of measurement. This
result leads to a locally changing refractive index which is an important point
and has to be emphasized. It is a qualitative difference to previously
existing self-consistent theories. The symmetry of mesoscopic transport is
broken by the boundary condition, and the frequency dependent absorption in
Silicon leads to two qualitatively differing mode types. Strong
confinement of a mode is in principle only possible when the symmetry in some
dimension is unbroken.\\ 
A different situation leads inevitably to extended
modes with diverse coherence properties. The computed correlation lengths $\xi$
and intensity distributions of both types of modes can be found in
Fig.\,\ref{DUAL}. The correlation length is a measure for the mode size, which can
be observed in the experiment.\\
In our calculations homogeneous dissipation for both mode types determined by the surface properties of the
disordered sample is assumed to be less than $10\%$ of the loss value through
the same area of boundary. The boundaries are as such symmetric and their
absorption can be found in \cite{Palik}. The symmetric mode Fig.\,\ref{DUAL}\,(b) suffers only
homogeneous loss and is strongly confined due to self-consistent photon propagation throughout the system and obeys the
dissipation induced length $ \zeta $. The stationary state diameter (derived at the decay of 1/e)
 in this case is of $2.2\times l_s$. Definitely completely different behaves
the extended mode caused by spatial symmetry breaking Fig.\,\ref{DUAL}\,(c),
which consequently also results in a partial break of the time-reversal
symmetry. The mode therfore obeys the dissipative length scale $\chi_{d}$. Both lengths are deduced
within  Eq.\,\ref{Pole}. The denominator carries an inherent differential
with respect to the limited dimension due to the broken symmetry through loss. With respect to the finite
dimension extended modes cover the whole extent of the sample. The mode is elliptically shaped, and the correlation length at the
interface to the substrate is by far reduced when compared to the center.\\
A comparison of emitted intensity in spatial resolution Fig.\,\ref{DUAL}\,(a)
clearly shows that the extended laser mode obeys different laws than the
confined mode does. The difference is obvious when comparing both profiles,
the restricted y-dimension (blue dashed graph) and the wide x-dimension (solid
lines). The aspect ratio is clearly of fundamental importance even though
both directions are by far longer than the scattering mean free path which is
in both cases about $l_s= 500\,nm$ (details see caption of
Fig.\,\ref{DUAL}). $l_s$ is very short due to high self-consistent
non-linearities. These non-linearities lead to a significantly different
refractive index for pumped material compared to a result derived by CBS far
below the threshold. The behavior can be explained in our
model of coherent intensity propagation that differs from the so-called
quasi-mode model. In contrast to quasi-modes, that are deduced from a
diagrammatic {\it single particle picture} of the electromagnetic wave,
our results go much further. The correlation length $\xi$ can be interpreted
as a measure of the
mode in the {\it correlated two
particle picture}, the coherence of the wave and simultaneously the coherent
transport of intensity. It can be explained as follows: Especially the
interference contributions (Cooperons) suffer from dissipation, meaning the
symmetry break in-plane especially reduces interference effects. Consequently
the degree of coherence of transported intensity in the situation
Fig.\,\ref{DUAL}\,(b) below the threshold is much higher - apart from all
spatial effects - than in situation Fig.\,\ref{DUAL}\,(c). This transport
inherent coherence leads to a stronger accumulation of intensity but even more
to higher non-linear gain. The concrete meaning is that the unlimited
development of the Cooperon drives the system very fast to the lasing
transition, which as such guarantees energy conservation in stationary
state. Lethokov's bomb argument in other words is avoided because the system
prefers to lase.\\
In case of extended modes the situation changes. Cooperons
are inhibited in their development due to the loss to the boundaries. Finally
it can be deduced that extended modes are preferentially built up by
incoherent contributions and the accumulated intensity is locally renormalized
by the loss through the boundary. In both cases Cooperon
  contributions remain to be the coherent stimulation process of emission.\\
 This positional distribution of the modes'
coherent intensities are displayed in the color coding of Fig.\,\ref{DUAL}\,(b)
and (c). Corresponding laser thresholds to both mode types can be compared in Fig.\,\ref{thresh}. It can be clearly seen that the confined mode (green) reaches the
threshold for by far lower pumping intensity $\gamma_p$ than the extended mode
(blue) does.\\
It is noted that this approach is in principle usable to
   construct spatially overlapping as well as spectrally coupled laser
   switches \cite{Leonetti1}, the mode coupling can be by far more
   sophisticatedly designed than in this paper.\\

\section{Conclusion} The solution of a complicated statistical behavior like
that of random lasing in granular disordered matter visualizes the power of Vollhardt-W\"olfle transport theory
of photons which we coupled to laser rate equations self-consistently. 
The correlation length for the intensity at the laser
transition derived by diagrammatic transport theory describes the
modes' shapes and diameters and it includes the conditions of spatially uniform
as well as time-reversal symmetry breaking losses. Engineering highly frequency selective
substrates is an efficient tuning mechanism for spectrally very close modes
which arise at different thresholds and arrive at completely different shapes
and diameters due to the breaking of the spatial symmetry. Confined modes
exist due to unbroken spatial symmetry, extended modes arise due to spatially
non-uniform position-dependent losses and non-linearities. The conception of the mode we derive
in this work is fundamentally different from the quasi-mode picture, which is
a single-particle picture results. Our theory defines the lasing modes as
correlations between scattered photons. The here presented results for lasing
mode diameters, are inseparable from the two-particle picture and they are a
measure of coherent transported intensity in granular amplifying media at the
laser threshold. A breaking of positional symmetry leads to the formation of
extended modes and in the same process to their pinning. It has to be pointed
out that the aspect ratio of
the sample is relevant for the modes' shapes, even though the samples are by
far larger in every extent than the scattering mean free path. Additionally a symmetry break due to loss leads to significantly differing lasing threshold
behavior, varying gain and gain-saturation dynamics. We hope that these results stimulate further research,
theoretically and experimentally, on the modal behavior of random lasers.\\

{\bf Acknowledgments} The authors thank B. Shapiro, K. Busch, J. Kroha, C. M. Soukoulis, H. Cao, A. D. Stone,
O. Muskens, A. Lagendijk and B. A. van Tiggelen for highly valuable discussions. H. Kalt and his
group are
gratefully acknowledged for providing insights to the world of
experiments. Support by the Deutsche Forschungsgemeinschaft (DFG) through
project GSC21
Karlsruhe School of Optics \& Photonics (KSOP) is acknowledged.\\


\vspace*{1.0cm}

\begin{thebibliography}{99999999999}


\bibitem{Cao} H. Cao, J. Y. Xu, D.Z Zhang, S. H. Chang, S. T. Ho,
  E. W. Seelig, X. Liu, R. P. H. Chang, ``Spatial Confinement of Laser Light in Active Random Media.'',
{\em Phys. Rev. Lett.} {\bf 84}, 5584 (2000).

\bibitem{Vaneste01}
C. Vanneste, P. Sebbah, "Selective excitation of localized modes in active random media", 
{\em Phys. Rev. Lett.} {\bf 87}, 183903 (2001).

\bibitem{Mujumdar04}
S. Mujumdar, M. Ricci, R. Torre, D. S. Wiersma, ``Amplified Extended Modes in Random Lasers'',
{\em Phys. Rev. Lett.} {\bf 93}, 5, 053903 (2004).

\bibitem{Vaneste07}
C. Vanneste, P. Sebbah and H. Cao, "Lasing with resonant feedback in weakly scattering random systems", 
{\em Phys. Rev. Lett.} {\bf 98}, 143902 (2007).

\bibitem{Muskens}
R. G. S. El-Dardiry, A. P. Mosk, O. Muskens, A. Lagendijk, ``Experimental studies on the mode structure of random lasers'',
{\em Phys. Rev. A} {\bf 81}, 043830 (2010).

\bibitem{Jaq}
P. Stano, P. Jacquod, `` Overlapping lasing modes in the Anderson localized
regime'', {\em Nature Photonics} {\bf 7}, 66-71 (2013).

\bibitem{Stone}
H. E. T\"ureci, L. Ge, S. Rotter, A. D. Stone, `` Strong Interactions in
Multimode Random Lasers'', {\em Science} {\bf 320}, 643 (2008).


\bibitem{Leonetti}
M. Leonetti, C. Conti, C. Lopez,``The mode-locking transition of random lasers'',
{\em Nature Photonics} {\bf 5}, 615 (2011).

\bibitem{Mosk_Mode}
K. L. van der Molen, R. W. Tjerkstra, A. P. Mosk, A. Lagendijk, ``Spatial Extent of Random Laser Modes'',
{\em Phys. Rev. Lett.} {\bf 98}, 14901 (2007).

\bibitem{Vardeny}
R. C. Polson, Z. V. Vardeny, ``Spatially mapping random lasing cavities'',
{\em Optics Letters}, {\bf 35}, 16, 2801 (2010).

\bibitem{Kalt09}
J. Fallert, R. J. B. Dietz, J.  Sartor, D. Schneider, C. Klingshirn, H. Kalt ``Co-existence of strongly and weakly localized random laser modes'',
{\em Nature Photonics} {\bf 3}, 279 (2009).

\bibitem{Kalt10}
 H. Kalt, J. Fallert, R. J. B. Dietz, J. Sartor, D. Schneider, C. Klingshirn, ``Random lasing in nanocrystalline ZnO powders'',
{\em phys. stat. sol.} (b) {\bf 247}, 1448 (2010).

\bibitem{Woelfle80}
D. Vollhardt, P. W\"olfle,
``Diagrammatic, self-consistent treatment of the Anderson  localization problem in $d <= 2$ dimensions.'',
{\em Phys. Rev. B} {\bf 22}, 4666 (1980).

\bibitem{Akkermans}
E. Akkermans, R. Maynard,
``Weak localization of waves'', {\em J. Physique Lett.} {\bf 46}, 1045 (1985).

\bibitem{Yamilov}
A. G. Yamilov, R. Sarma, B. Redding, B. Payne, H. Noh, H. Cao,
``Position-dependent diffusion of light in disordered waveguides'', {\em
  Phys. Rev. Lett.} {\bf 112}, 023904 (2014).

\bibitem{Shapiro}
V. M. Apalkov, M. E. Raikh, B. Shapiro, ``Random resonators and prelocalized
modes in disordered dielectric films'', {\em Phys. Rev. Lett.} {\bf 89},
016802 (2002).

\bibitem{Andreasen}
J. Andreasen, H. Cao, ``Numerical study of amplified spontaneous emission and lasing in random media'',
{\em Phys. Rev. A} {\bf 82}, 063835 (2010).

\bibitem{Frank11}
R. Frank, A. Lubatsch,
``Scalar wave propagation in random amplifying media: Influence of localization effects on length and time scales and threshold behavior'',
Phys. Rev. A {\bf 84}, 013814 (2011).

\bibitem{Lubatsch05}
A. Lubatsch, J. Kroha, K. Busch,``Theory of light diffusion in disordered media with linear absorption or gain'',
{\em Phys. Rev. B} {\bf 71}, 184201 (2005).

\bibitem{Maret}
G. Maret, T. Sperling, W. Buehrer, A. Lubatsch, R. Frank, C. M. Aegerter,
``Reply to comment by F. Scheffold and D. Wiersma: Inelastic scattering puts in question recent claims of Anderson
localization of light'', {\em Nature Photonics} {\bf 7}, 934–935 (2013).

\bibitem{Palik}
E. D. Palik, ``Handbook of Optical Constants'', Elsevier (1997).

\bibitem{Mie}
G. Mie,``Beitr\"age zur Optik tr\"uber Medien, speziell kolloidaler
Metall\"osungen'', {\em Annalen  der Physik }, {\bf 4} , 25, 377-445 (1908).

\bibitem{Lagendijk06}
K. L. van der Molen,  P. Zijlstra, A. Lagendijk, A. P. Mosk , ``Laser threshold of Mie resonances'',
{\em  Optics Letters} {\bf 31}, 1432 (2006).

\bibitem{John04}
L. Florescu, S. John,
``Lasing in a random amplifying medium: Spatiotemporal characteristics and nonadiabatic atomic dynamics''
{\em Phys. Rev. E} {\bf 70}, 036607 (2004).

\bibitem{Leonetti1}
M. Leonetti, C. Conti, C. Lopez, ``Tunable degree of localization in random lasers with controlled interaction'', Appl. Phys. Lett. 101, 051104 (2012).
\end{thebibliography}
\end{document}